\documentclass{article}
\usepackage{hiph-art}
\usepackage{graphicx}
\usepackage{epsfig}
\volnumber{25} \issuenumber{1} \edyear{2006}                             
\frompage{000} \topage{000}                                              
\recrevdate{7 May 2005}                                                  
\def\rightmark{Multiparticle production...}
\def\leftmark{O.V.Utyuzh et al.}

\title{Multiparticle production processes\\
from the Information Theory point of view}
\authors{
{O.V.Utyuzh$^{1,a}$, G.Wilk$^{1,b}$ and Z.W\l odarczyk$^{2}$ %
\index{One, O.V.Utyuzh } 
\index{Two, G.Wilk} 
\index{Three, Z.W\l odarczyk}
}\\[2.812mm]
{\normalsize
\hspace*{-8pt}$^1$ The Andrzej So\l tan Institute for Nuclear Studies;\\
Ho\.za 69; 00-681 Warsaw, Poland\\[0.2ex]
\hspace*{-8pt}$^2$ Institute of Physics, \'Swi\c{e}tokrzyska Academy,\\
         \'Swi\c{e}tokrzyska 15; 25-406 Kielce, Poland; E-mail:
         wlod@pu.kielce.pl\\[0.2ex]
\hspace*{-8pt}$^a$ E-mail: utyuzh@fuw.edu.pl\\[0.2ex]
\hspace*{-8pt}$^b$ E-mail: wilk@fuw.edu.pl
}}

\abstract{We look at multiparticle production processes from the Information Theory
point of view, both in its extensive and nonextensive versions. Examples of both
symmetric (like $pp$ or $AA$) and asymmetric (like $pA$) collisions are considered
showing that some ways of description of experimental data used in the literature are
of more general validity than usually anticipated.}



\makeindex
\begin{document}

\maketitle

\section*{Introduction}\label{intro}

The multiparticle production processes are usually first approached
by means of statistical models \cite{Stat} in order to make quick
estimations of such parameters as temperature $T$ or chemical
potential $\mu$ of the hadronizing matter (with tacit assumptions
that $T$ and $\mu$ have the usual meaning in the realm of hadronic
production). It means that "thermal-like" (i.e. exponential) form of
relevant distribution (in transverse momentum $p_T$ or in rapidity
$y$) is used to fit data,
\begin{equation}
f(X) \sim \exp\left[ - X/\Lambda \right] . \label{eq:EXP}
\end{equation}
This formula occurred to be very robust, mainly because (cf.,
\cite{LVH}) the $N-1$ unmeasured particles act as a {\it heath bath}
which action on the observed particle is described by a single
parameter $\Lambda$ identified (for system in thermal equilibrium)
with temperature $T$. Discrepancies from (\ref{eq:EXP}) observed in
many places are then attributed to the fact that in reality such a
"heath bath" is neither infinite nor homogeneous and one needs more
parameters for its description. The minimal extension would be to
regard $\Lambda$ as being $X$-dependent, for example, $\Lambda
\simeq \Lambda_0 + a\cdot X$ \cite{Alm}, what results in
\begin{equation}
f_q(X) \sim \exp_q\left[ - X/\Lambda_0 \right] \stackrel{def}{=}
     \left[1 - (1-q)X/\Lambda_0\right]^{1/(1-q)}
     \qquad {\rm where}\qquad q=1 + a .
        \label{eq:EXPq}
\end{equation}
In the case when $X$ is energy $E$ and $\Lambda$ is temperature $T$
the coefficient $a$ is just the inverse heat capacity, $a=1/C_V$
\cite{Alm}. Such phenomenon leads to the so-called {\it
nonextensivity} with $q$ being the {\it nonextensivity parameter}.
Notice that for $q\rightarrow 1$ (or for $a\rightarrow 0$) $f_q(X)$
becomes $f(X)$. We show here that eqs. (\ref{eq:EXP}) and
(\ref{eq:EXPq}) originate in a natural way from information theory
({\it IT}) approach in its, respectively, extensive (based on
Shannon entropy) and nonextensive (based on Tsallis entropy) forms.
In what follows (Section \ref{IT}) we shall briefly present this
approach illustrating it (in Sections \ref{RESYM} and \ref{REASYM})
with some fits to the existing data.

\section*{Information Theory and Multiparticle Production Processes}\label{IT}

To introduce IT in the present context let us consider typical
situation: experimentalists obtain some new and intriguing data.
Immediately these data become subject of interest to theoreticians
and in no time a number of distinctive and unique (as concerns
assumptions and physical concepts) explanations is presented, which,
disagreeing about physical concepts used, nevertheless all fit these
data. The natural question arises: {\it which of the proposed models
is the right one?} The answer is: {\it all of them} (to some
extent). This is because experimental data are providing only
limited amount of information and all models mentioned here are able
to reproduce it. To quantify this reasoning one has to define, using
{\it IT}$^a$, the notion of information. Its extensive version is
based on the Shannon information entropy,
\begin{equation}
S = - \sum_i \, p_i\, \ln p_i \, , \label{eq:Shannon}
\end{equation}
where $p_i$ denotes probability distribution of interest. The {\it
least possible information}, corresponding to equal probability
distribution of $N$ states, $p_i = 1/N$, results in {\it maximal
entropy}, $S=\ln N$. The opposite situation of {\it maximal
information}, when only one state is relevant (i.e. $p_l=1$ and
$p_{i\neq l}=0$) results in {\it minimal entropy}, $S=0$. Denoting
by $\langle R_k\rangle$ {\it a priori} information available on
experiment, like conservation laws and results of measurements of
some quantities $R_k$, one is thus seeking probability distribution
$\{p_i\}$ such that: $i)$ it tells us {\it the truth, the whole
truth} about our experiment, i.e. in addition to being normalized it
reproduces the known results:
\begin{equation}
\sum_i p_i =1\qquad {\rm and}\qquad \sum_i\, p_i\, R_k(x_i)\, =\,
\langle R_k\rangle\, , \label{eq:constraints}
\end{equation}
and $ii)$ it tells us {\it nothing but the truth} about our
experiment, i.e. it conveys {\it the least information} (only the
information contained in this experiment).

To find such $\{p_i\}$ one has to {\it maximize} the Shannon entropy
under the above constraints (therefore this approach is also known
as MaxEnt method). The resultant distribution has familiar
exponential shape
\begin{equation}
p_i\, =\, \frac{1}{Z}\, \cdot \exp\left[ - \sum_k\lambda_k\cdot R_k(x_i) \right]\, .
\label{eq:MEp}
\end{equation}
Although it looks identical to the "thermal-like" (Boltzmann-Gibbs)
formula (\ref{eq:EXP}) there are {\it no free parameters} here
because $Z$ is just normalization constant assuring that $\sum p_i
=1$ and $\lambda_k$ are the corresponding lagrange multipliers to be
calculated from the constraint equations (\ref{eq:constraints}).$^b$

It is worth to mention at this point \cite{MaxEnt} that the most
probable multiplicity distribution $P(n)$ in the case when we know
only the mean multiplicity $\langle n\rangle$ of distinguishable
particles is geometrical distribution $P(n) = \langle
n\rangle^n/(1+\langle n\rangle)^{(n+1)}$ (which is broad in the
sense that its dispersion is $D(n) \sim \langle n\rangle$).
Additional knowledge that all these particles are indistinguishable
converts the above $P(n)$ into Poissonian form, $P(n) = \langle
n\rangle^n\exp ( -\langle n\rangle )/n!$ which is the narrow one in
the sense that now its dispersion is $D(n) \sim \sqrt{\langle
n\rangle}$. In between is situation in which we know that particles
are grouped in $k$ equally strongly emitting sources, in which case
one gets Negative Binomial distribution \cite{NBin}$^c$
$$P(n) = \Gamma(k+n)/\left[\Gamma(n+1)\Gamma(k)\right]\cdot \left(
\frac{k}{\langle n\rangle }\right)^k/\left[ 1 + \frac{k}{\langle
n\rangle}\right]^{k+n}$$.

The other noticeable example provided in \cite{Chao} is the use of
{\it IT} to find the minimal set of assumptions needed to explain
all multiparticle production data of that time. They were equally
well described by models like multi-Regge, uncorrelated jet,
thermodynamical, hydrodynamical etc., which, after closer scrutiny,
turned out to share (in {\it explicit} or {\it implicit} manner),
two basic assumptions: $i)$ that only part $W=K\cdot\sqrt{s}$
($0<K<1$) of the initially allowed energy $\sqrt{s}$ is used for
production of observed secondaries (located mostly in the center
part of the phase space; in this way {\it inelasticity} $K$ found
its justification \cite{Kq}, it turns out that $K\sim 0.5$); $ii)$
that transverse momenta of produced particles are limited and the
resulting phase space is effectively one-dimensional (dominance of
the longitudinal phase space). All other assumptions specific for a
given model turned out to be spurious.$^d$

Suppose now  that some new data occur which disagree with the
previously established form of $\{p_i\}$. In {\it IT} approach it
simply signals that there is some additional information not yet
accounted for. This can be done either by adding a new constraint
(resulting in new $\lambda_{k+1}$, cf., for example \cite{Gauss}, we
shall not discuss it here) or by using some other form of {\it IT},
for example its nonextensive version ({\it IT}$_q$). The later is
necessary step for systems which experience long range correlations,
memory effects, which phase space has fractal character or which
exhibit some intrinsic dynamical fluctuations of the otherwise
intensive parameters (making them extensive ones, like $T$ here).
Such systems are truly {\it small} because the range of changes is
of the order of their size. In this case the Shannon entropy
(\ref{eq:Shannon}) is no longer a good measure of information and
should be replaced by some other measure. Out of infinitely many
possibilities \cite{ENT} we shall choose Tsallis entropy \cite{CT},
\begin{equation} S_q\,
=\, - \frac{1}{1-q}\sum_i\, \left( 1 - p_i^q \right)\quad \
                   \stackrel{q \rightarrow 1}{\Longrightarrow}\quad
                   S_{q=1}=  - \sum_i \, p_i\, \ln p_i ,\label{eq:Tsallis}
\end{equation}
which goes over to Shannon form (\ref{eq:Shannon}) for $q\rightarrow
1$. The $\{p^{(q)}_i\}$ are obtained in the same way as before but
with modified constraint equation:$^e$
\begin{equation}
\sum_i\, \left[p_i\right]^q\, R_k(x_i)\, =\, \langle R^{(q)}_k\rangle .
\label{eq:q_constr}
\end{equation}
This leads to formally the same formula for $p_i=p^{(q)}_i$ as in
Eq. (\ref{eq:MEp}) but with $Z \rightarrow Z_q$ and $\exp (...)
\rightarrow \exp_q (...)$. Because such entropy is nonextensive,
i.e., $S_{q(A+B)} = S_{qA}+S_{qB} + (1-q)S_{qA}\cdot S_{qB}$
\cite{CT}, the whole approach became known as nonextensive (Tsallis)
statistics. It should be stressed at this point that the
nonextensivity parameter $q$ cumulates action of all possible
dynamical sources causing deviation from the usual Boltzmann-Gibbs
statistics or Shannon entropy and as such can be considered as a
useful phenomenological parameter allowing to describe data without
deciding which of dynamical models leading to the same $q$ is the
right one \cite{Biya}. However, in what follows we shall mainly
address one possible sources of $q\neq 1$, namely intrinsic
fluctuations present in the system represented by fluctuations of
the $1/\Lambda$ parameter. This is because, as shown in \cite{WW},
in this case
\begin{equation}
q\, =\, 1\, \pm\,
         \left[
             \left\langle \left(1/\Lambda\right)^2\right\rangle\,
             -\,
             \left\langle 1/\Lambda \right\rangle^2
        \right] /
             \left\langle 1/\Lambda \right\rangle^2 , \label{eq:Q}
\end{equation}
i.e. parameter $q$ is a measure of such fluctuations (with $\langle
\dots\rangle$ denoting the respective averages over them.$^f$)

\section*{Confrontation with Experimental Data - Symmetric Case}\label{RESYM}

We shall now confront these ideas with reality. At first we shall
remind shortly main results of our recent investigations of
hadronizations taking place in collisions of symmetric systems like
pp and $\rm p\bar{p}$ \cite{Kq,NuCim,Cern}, heavy ions $AA$
\cite{MaxEntq,Trends,JPG} or $\rm e^+e^-$ \cite{Nukleon}. This will
be followed by some new results on collisions of asymmetric systems
exemplified by ${\rm p}A$ collisions. It must be stressed that what
we are proposing is not a new model but rather sets of {\it least
biased, most probable} distributions describing data by accounting
for available information provided in terms of constraints emerging
from conservation laws and from some previously known experimental
facts (or from some assumed dynamical input, which is thus
confronted with experimental data).

Hadronization means that some invariant energy $M$ (assumed to be
known) gives rise to a number $N$ of secondaries (also assumed to be
known) and question asked is: in what way these secondaries are
distributed in the allowed phase space? So far distributions in
transverse momenta $p_T$ were treated separately \cite{MaxEntq,JPG}
from distributions in rapidity \cite{MaxEntq,Kq,NuCim,Cern,Trends},
for which it was always assumed that mean transverse mass
$\mu_T=\sqrt{m^2 +\langle p_T\rangle^2}$ was known, i.e. that we are
working in the effectively $1$-dimensional phase space.

\begin{figure}[ht]
  \begin{minipage}[ht]{57mm}
        \insertplot{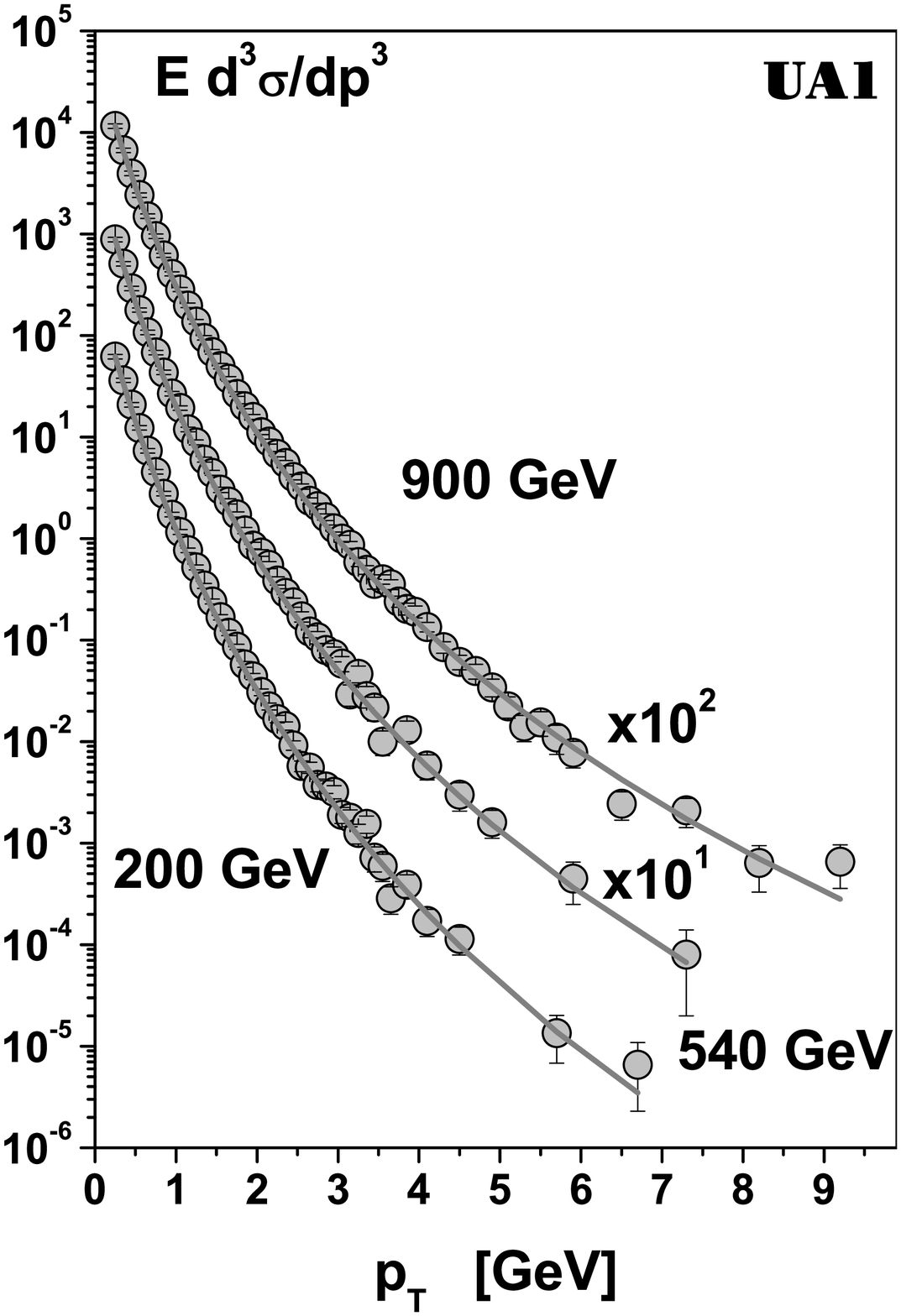}
  \end{minipage}
\hspace{5mm}
  \begin{minipage}[ht]{57mm}
        \insertplot{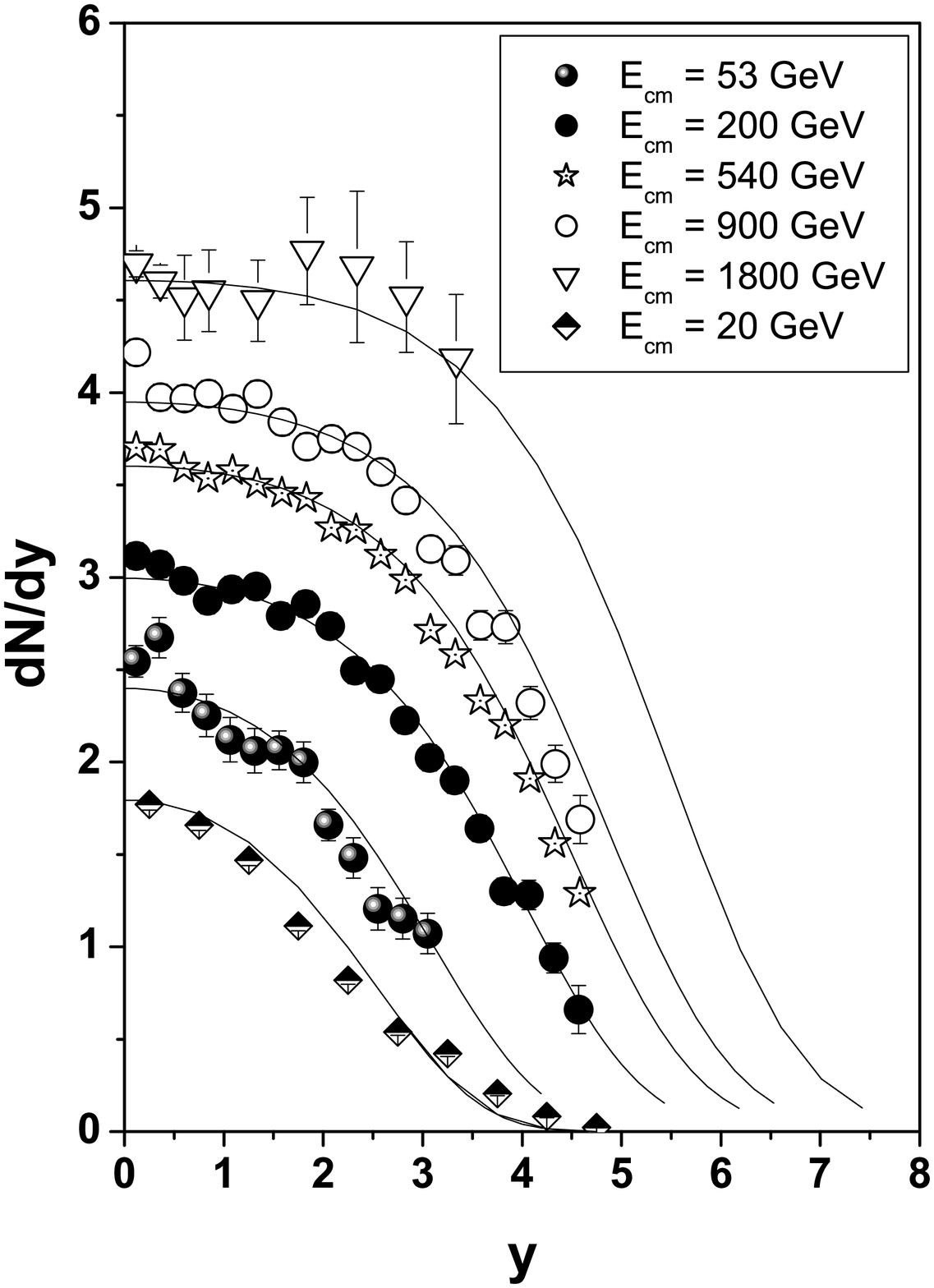}
  \end{minipage}
  \vspace{-3mm}
  \caption{ \footnotesize {Example of description of data on $p_T$ spectra
            from UA1 experiment on ${\rm p\bar{p}}$ collisions \protect\cite{MaxEntq}
            (left panel) and data on rapidity spectra in $\rm pp$ and ${\rm p\bar{p}}$
            collisions \protect\cite{Kq} (right panel).}}
 \label{fig:Fig1}
\end{figure}

Concerning $p_T$, distributions it has been shown that using
$q$-statistics (with $q\simeq 1.05$ for $AA$ collisions \cite{JPG}
to $q\simeq 1.1$ for $\rm p\bar{p}$ collisions \cite{MaxEntq}) one
can describe data in large domain of $p_T$, see, for example, Fig.
\ref{fig:Fig1}. The characteristic feature is that now values of $q$
are much smaller than those obtained fitting data in longitudinal
phase space (cf. \cite{MaxEntq} for discussion). Following
\cite{WW,JPG} (see also \cite{Alm,CB,Biro}) we argue that $q>1$
shows existence of the fluctuations of temperature in the
hadronizing system mentioned at the beginning. In fact, the
$q=1.015$ for $AA$ collisions \cite{JPG} corresponds (according to
(\ref{eq:Q})) to fluctuation of $T$ of the order $\Delta T/T \simeq
0.12$, which do not vanish with increasing multiplicity \cite{WW}.
These fluctuations exist in small parts of the hadronic system with
respect to the whole system, they are not of the event-by-event
type, for which $\Delta T/T \sim 0.06/\sqrt{N} \rightarrow 0$ for
large $N$. It should be stressed at this point that such
fluctuations are very interesting \cite{fluc} because they provide a
direct measure of the total heat capacity of the system, $C$:
\begin{equation}
\frac{\sigma^2(\beta)}{\langle \beta\rangle^2} \stackrel{def}{=} \frac{1}{C} = q - 1.
\label{eq:fluct}
\end{equation}
In fact, because $C$ grows with the collision volume $V$ of reaction
we expect that $q({\rm hadronic}) > q({\rm nuclear})$ which seems to
be indeed observed (cf. also \cite{Biya} where nuclear collisions
data at different centralities providing direct access to volume $V$
were analyzed).

When applied to rapidity distributions {\it IT} method leads to formula formally
identical with formulas used in statistical models (cf., for example, like
\cite{CY}),
\begin{equation}
f_N(y)\, =\, \frac{1}{Z}\cdot \exp\left[ - \beta\cdot \cosh y \right]\quad {\rm
where}\quad Z = \int^{Y_M}_{-Y_M}\!\!\!dy \exp\left[ - \beta\cdot \cosh y \right].
\label{eq:ITstat}
\end{equation}
However, whereas in \cite{CY} (and in other similar models) $1/Z$
and $\beta$ were just two free parameters, here $Z$ is normalization
constant and $\beta=\beta (M,N,\mu_T)$ is obtained by solving
constraint equation,$^g$
\begin{equation}
\int^{Y_M}_{-Y_M}\!\!\!f_N(y) =1\qquad {\rm and} \qquad \int^{Y_M}_{-Y_M}\!\!\! dy
\left[ \mu_T\cdot \cosh y\right]\cdot f_N(y) = \frac{M}{N}. \label{eq:constr}
\label{eq:constry}
\end{equation}
It means that the parameter $\beta$ (inverse of the so-called {\it
partition temperature} introduced in \cite{CY} type of models) is
connected (via {\it IT} method used) with the {\it dynamical input}
given by: the allowed energy $M$ (usually taken as a fraction,
$M=K\sqrt{s}$, of the total energy of reaction with parameter $K\in
(0,1)$ being the so called {\it inelasticity} of the reaction),
number of secondaries produced $N$ and the mean transverse mass
$\mu_T$. In asymmetric collisions to be discussed in the next
section one would have to add also constraint imposed by momentum
conservation (which is satisfied automatically in the case of
symmetric collisions discussed at the moment).

As detailed in \cite{MaxEnt,Kq}, $f_N(y)$ changes from
$f_{N=2}(y)\!\! =\!\! \frac{1}{2}\!\left[
\delta(y\!-\!Y_M)\!\!+\!\!\delta(y\!+\!Y_M)\right]$ (with
$\beta(N\rightarrow 2)\rightarrow - \infty$) via $f_{N_0}={\rm
const}.$ (with $\beta (N_0\simeq \ln(M/\mu_T)) \simeq 0$) to
$f_{N\rightarrow N_{max}} = \delta(y)$ (with $\beta
(N_{max}=M/\mu_T) = + \infty$). In other words, for small
multiplicities $\beta$ is {\it negative} (a feature alien to any
statistical model!) and it becomes zero only for $N\rightarrow N_0$.
At this multiplicity $f_N(y)={\rm const}$., behaviour known as {\it
Feynman scaling}.$^h$ Its occurence means that energy dependence of
multiplicity follows that of the longitudinal phase space. For
$N>N_0$ additional particles have to be located near the middle of
phase space in order to minimize energy cost of their production. As
result $\beta
> 0$ now, in fact (see \cite{MaxEnt} for details) for some ranges of $\langle E\rangle=M/N$
quantity $\bar{\beta} = \beta \langle E\rangle$ remains
approximately constant. Needless to say that for $N\rightarrow
N_{max}$ {\it all particles} have to stay as much as possible at the
center and therefore $\beta(M,N\rightarrow N_{max}) \rightarrow
+\infty$, whereas $f_{N\rightarrow N_{max}}(y) \rightarrow
\delta(y)$.

For nonextensive approach $\exp(...)$ must be replaced by
$\exp_q(...)_q$ and  $N_0 \rightarrow N^{(q)}_0 \simeq \left( 2 \ln
N_{max}\right)^q$. As shown in \cite{Kq} $q$ acts now as a free
parameter allowing for changing separately the shape and the height
of $f^{(q)}_N$ (which were interlocked for $q=1$ case). For $q>1$
one enhances tails of distribution reducing at the same time its
height. For $q<1$ the effect is opposite and additionally there is
kinematical condition, $1-(1-q)\beta_q\mu_T\cosh y \ge 0$, reducing
in this case the allowed phase space.$^i$

In \cite{Kq} we have used $f^{(q)}_N$ to fit ${\rm p\bar{p}}$ and
$\rm pp$ data introducing inelasticity $K$ as a free parameter to be
deduced from data (cf. Fig. \ref{fig:Fig1}). We shall not discuss
here the inelasticity issue (see \cite{Kq} for details and further
references) but concentrate on the $q$ parameter, which turned out
to follow {\it the same} energy dependence as the experimentally
deduced parameter $1/k$ of the Negative Binomial (NB) multiplicity
distribution \cite{NB}. This finding prompted us to argue that
parameter $q$ in this case is accounting for a new bit of
information so far unaccounted for, namely for fluctuations in the
multiplicity of produced secondaries (notice that in the constraint
equation for $\beta$ we were using in this case experimental values
of the corresponding {\it mean multiplicities} only). In fact, it is
known \cite{NBin} that NB can be obtained from Poisson distribution
once one allows for fluctuations in its mean value $\bar{n}$ of the
gamma distribution type, namely
\begin{equation}
P(n)\, =\, \int_0^{\infty} d\bar{n} \frac{e^{-\bar{n}}\bar{n}^n}{n!}\cdot
         \frac{\gamma^k \bar{n}^{k-1} e^{-\gamma \bar{n}}}{\Gamma (k)} =
   \frac{\Gamma(k+n)}{\Gamma (1+n) \Gamma (k)}\cdot
   \frac{\gamma^k}{(\gamma +1)^{k+n}} \label{eq:PNBD}
\end{equation}
where $$ \gamma = \frac{k}{\langle \bar{n}\rangle} \hspace{10mm}{\rm
and}\hspace{10mm} \frac{1}{k} = \frac{\sigma^2(n_{ch})}{\langle
n_{ch}\rangle^2}- \frac{1}{\langle n_{ch}\rangle}.$$

\begin{figure}[ht]
  \begin{minipage}[ht]{57mm}
        \insertplot{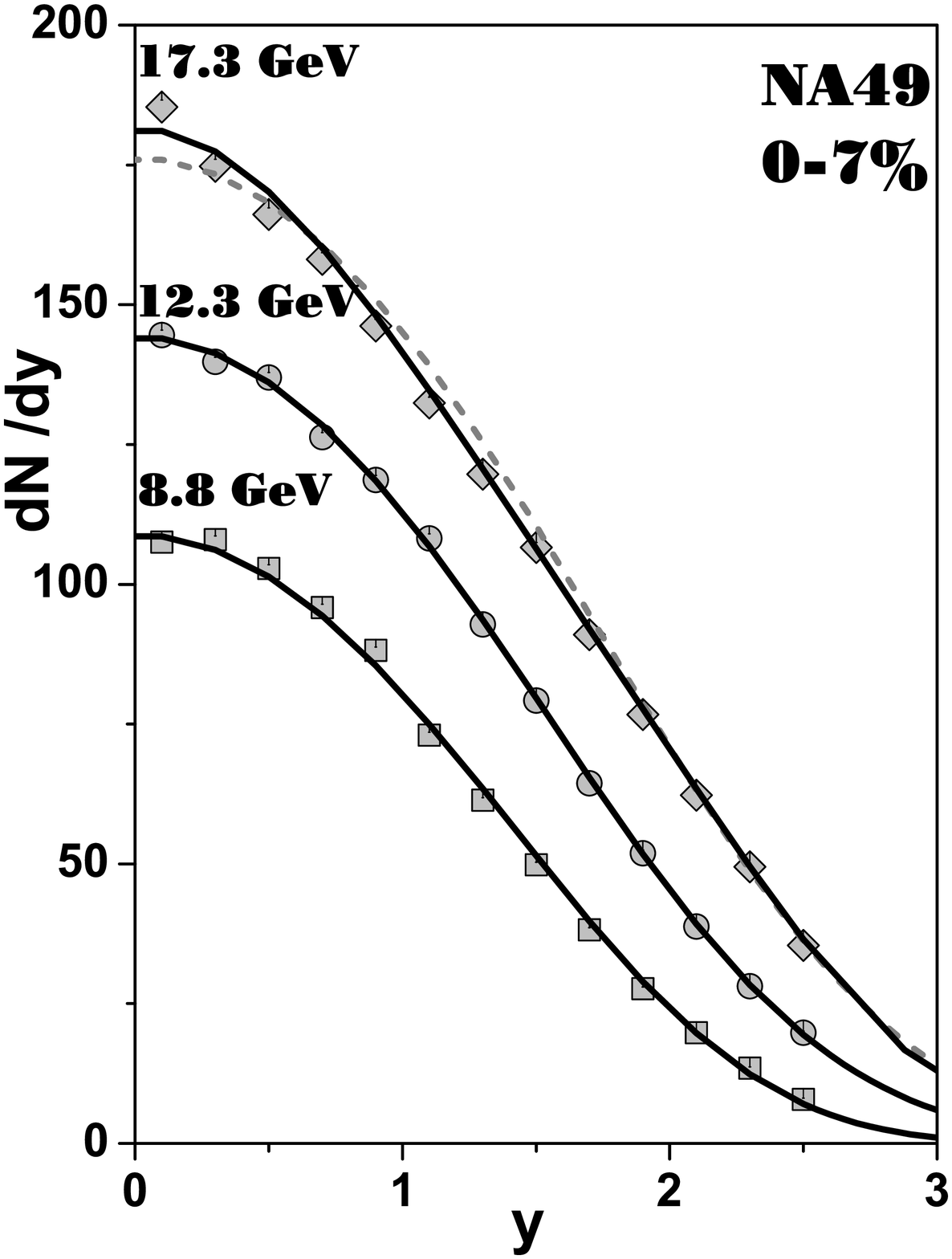}
  \end{minipage}
\hspace{5mm}
  \begin{minipage}[ht]{57mm}
        \insertplot{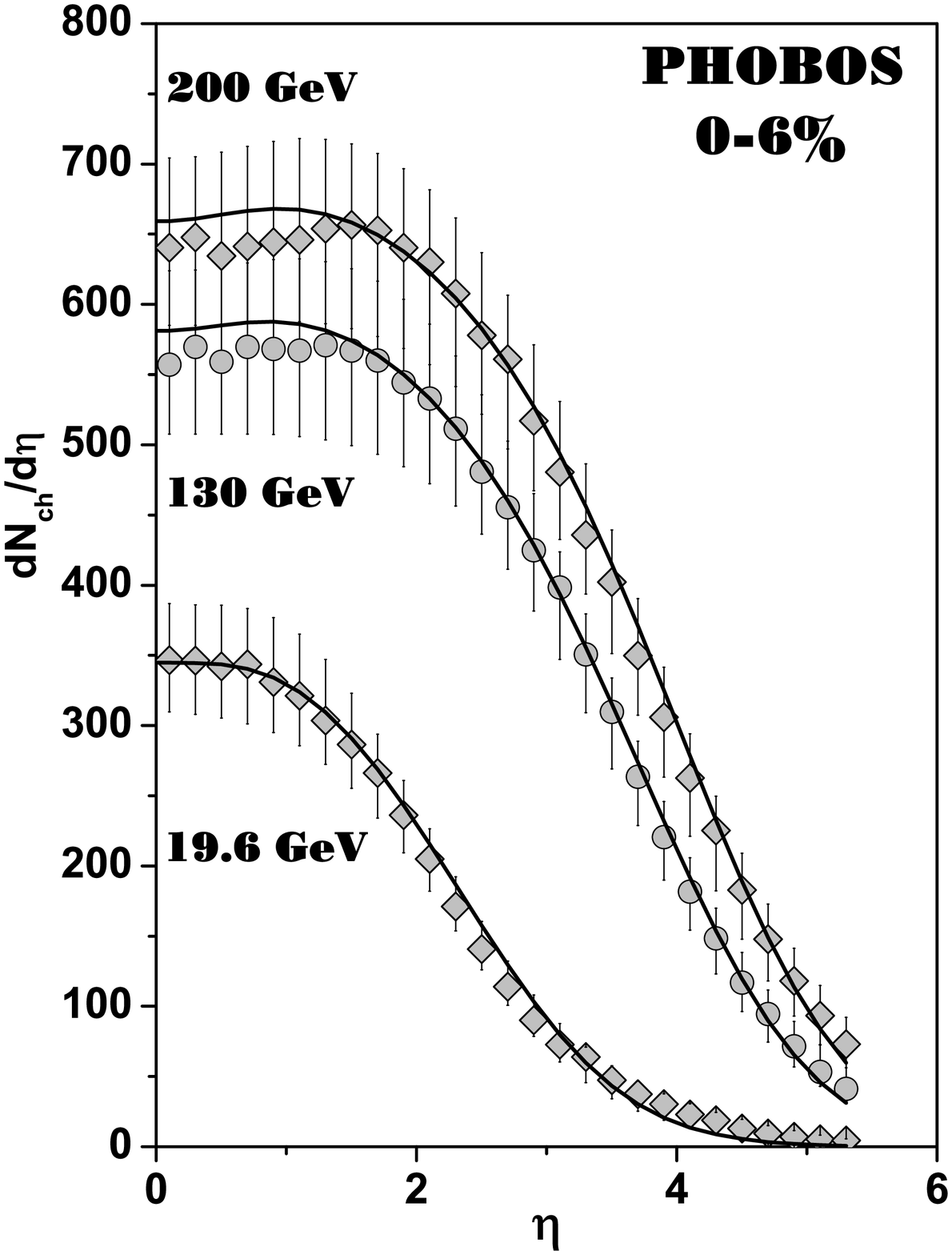}
  \end{minipage}
  \vspace{-3mm}
  \caption{ \footnotesize {Example of description of data on rapidity spectra
            from NA49 (left panel) for negatively charged pions produced in
            central $\rm PbPb$ collisions at different energies
            \protect\cite{MaxEntq}; the best fit for $17.3$ GeV is with
            additional (assumed) information of existence of two extensive
            sources located at $y =\pm \Delta y = 0.83$ in rapidity (see text for
            details). Right panel: fits to PHOBOS data for the most central
            ${\rm Au+Au}$ collisions \protect\cite{Trends}}}
 \label{fig:Fig2}
\end{figure}

Assuming now that these fluctuations contribute to nonextensivity
defined by the parameter $q$, i.e. that $D(\bar{n}) = q-1$ one gets
\begin{equation}
q\, =\, 1\, +\, \frac{1}{k}, \label{eq:qk}
\end{equation}
what we do observe (cf. also \cite{Cern}).

In Fig. \ref{fig:Fig2} we show comparison with $AA$ data at SPS
(NA49) \cite{MaxEntq} and RHIC (PHOBOS) \cite{Trends} energies. The
NA49 data can be fitted with $q=1.2$, $1.16$ and $1.04$ going from
top to bottom and so far we do not have plausible explanation for
these number. However, it should be stressed that at highest energy
two-component extensive source is preferable. The PHOBOS data with
$q=1.27$, $1.26$ and $1.29$ going from top to bottom. Extensive fits
do not work here at all. The first clear discepancy has been found
when fitting $e^+e^-$ data (see \cite{Nukleon}) where $dn/dy$ with
clear minimum for $y=0$ cannot be reproduced in such simple {\it
ITq} approach. It is thus obvious that there is some new information
we did not accounted for. It looks like we have two sources here
separated in rapidity (two jets of QCD analysis) but of no
statistical origin (rather connected with cascading process)
\cite{Nukleon}.

\section*{Confrontation with Experimental Data - Asymmetric Case}\label{REASYM}

Let us now proceed to a more complicated case of asymmetric
collisions exemplified by ${\rm p}A$ processes
\cite{NucData1,NucData2,LY}. In this case: $i)$ both colliding
objects are different and have different masses (one can therefore
expect that energy transfer to the central region from each of them
is not necessary the same as it was assumed before); $ii)$ the ${\rm
p}A$ collisions introduce a new element of uncertainty, the
effective size of the target, i.e. the number of nucleons, $\nu$,
from the nucleus $A$ with which the impinging nucleon is interacting
and the way this interaction proceeds.$^j$ The number $\nu$ can be
estimated from measurement of the number of "gray tracks"
\cite{gray}, therefore it will be included to our input information.
We shall analyze data \cite{NucData1} in which attempt was made to
isolate $dN_{\nu}/dy$ for $\nu =1,2,3,4$ and data \cite{NucData2}
where only the mean $\bar{\nu}$ number of collision is known
together with $dN_{\bar{\nu}}/dy$. The {\it a priori} available
invariant energy $\sqrt{s_{\nu}}$ in such a case is equal to (we
neglect all nuclear binding and Fermi motion effects): $ s_{\nu} =
\nu s + (\nu -1)^2 m^2$ where  $s=2m^2 + 2m\sqrt{p^2_{LAB} + m^2}$
is the invariant energy squared for $NN$ collisions. 
\begin{figure}[ht]
\begin{center}
\epsfig{file=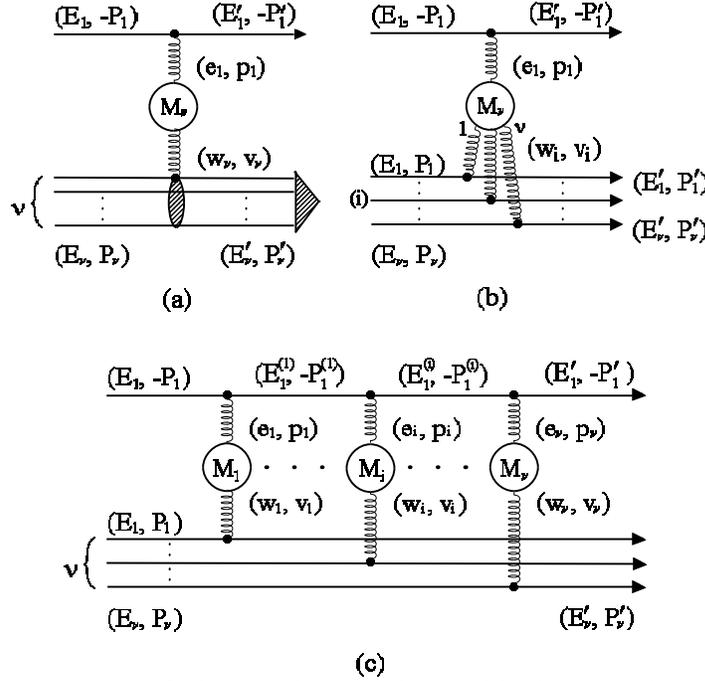, width=95mm}
\end{center}
\vspace{-5cm}
  \caption{\footnotesize Schematic views of tube $(a)$ and sequential $(c)$
           models of the $pA$ collisions. The case $(b)$ is essentially
           a particular realization of the tube type of model $(a)$.}
 \label{fig:Fig3}
\end{figure}

The assumed knowledge of $\nu$ must be supplemented by the known dependence of total
mean multiplicity of secondaries produced in the central region, $\bar{N}_{\nu}$, on
the mean number of inelastic collisions $\langle \nu\rangle$,
\cite{NucData1,NucData2}:
\begin{equation}
\bar{N}_{\nu} = \frac{1}{2}(1 + \langle \nu\rangle)\bar{N} \label{eq:multi}
\end{equation}
($\bar{N}$ is multiplicity of particles produced in $NN$ collisions
at the same energy $\sqrt{s_{\nu=1}}$).$^k$ To be able to apply {\it
IT} methods we must additionally decide whether ${\rm p}A$ collision
is more like a two body collision between a kind of "tube" of mass
$m_{\nu} = \nu m$ containing $\nu$ collectively acting nucleons and
single nucleon of mass $m$ (cf. Fig. \ref{fig:Fig3}(a) and (b)), or
whether it is rather a sequence of $\nu$ consecutive collisions of
the impinging nucleon with $\nu$ nuclear nucleons (cf. Fig.
\ref{fig:Fig3}c).

\begin{figure}[ht]
  \begin{minipage}[ht]{57mm}
        \insertplot{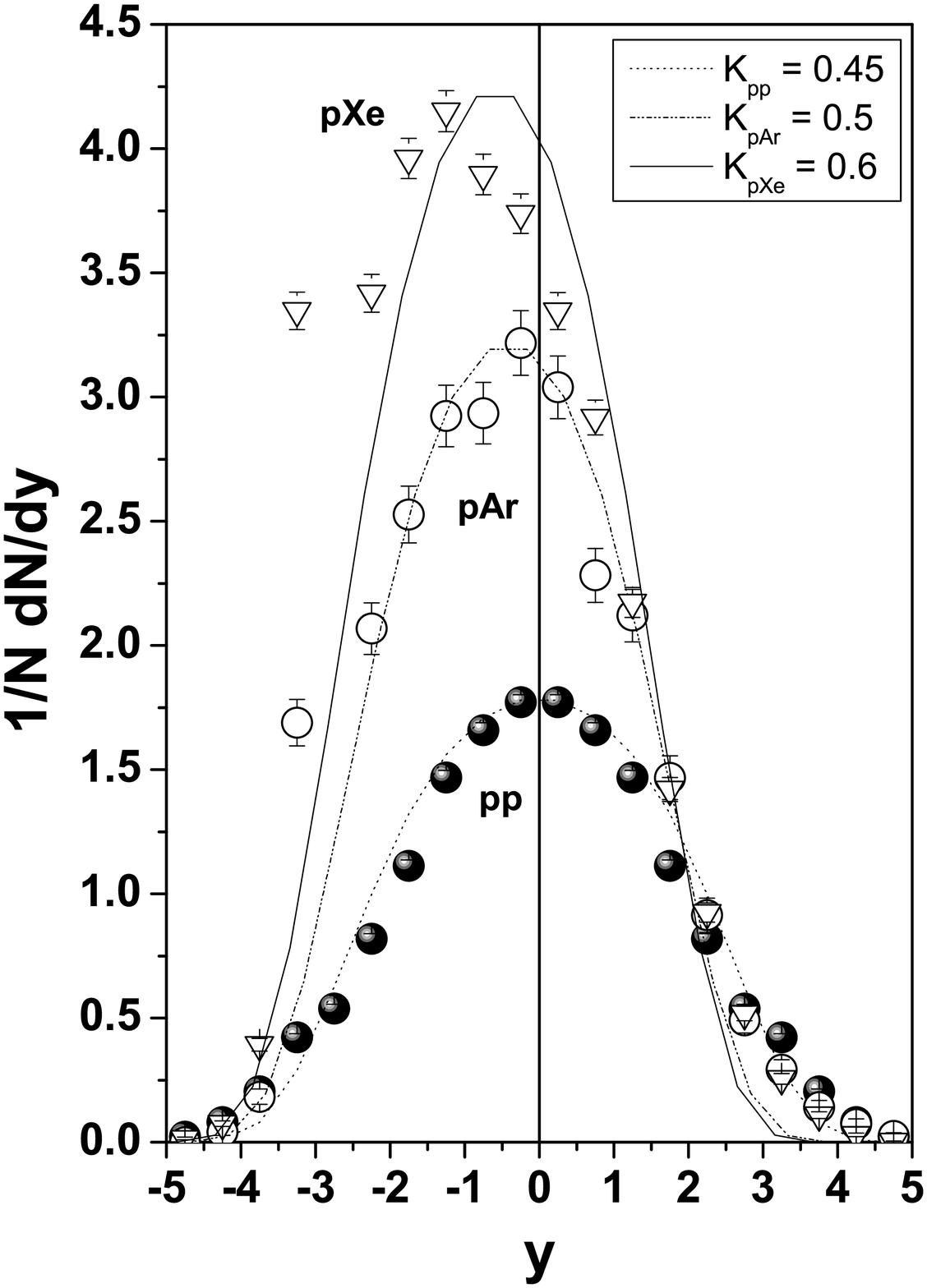}
  \end{minipage}
\hspace{5mm}
  \begin{minipage}[ht]{57mm}
        \insertplot{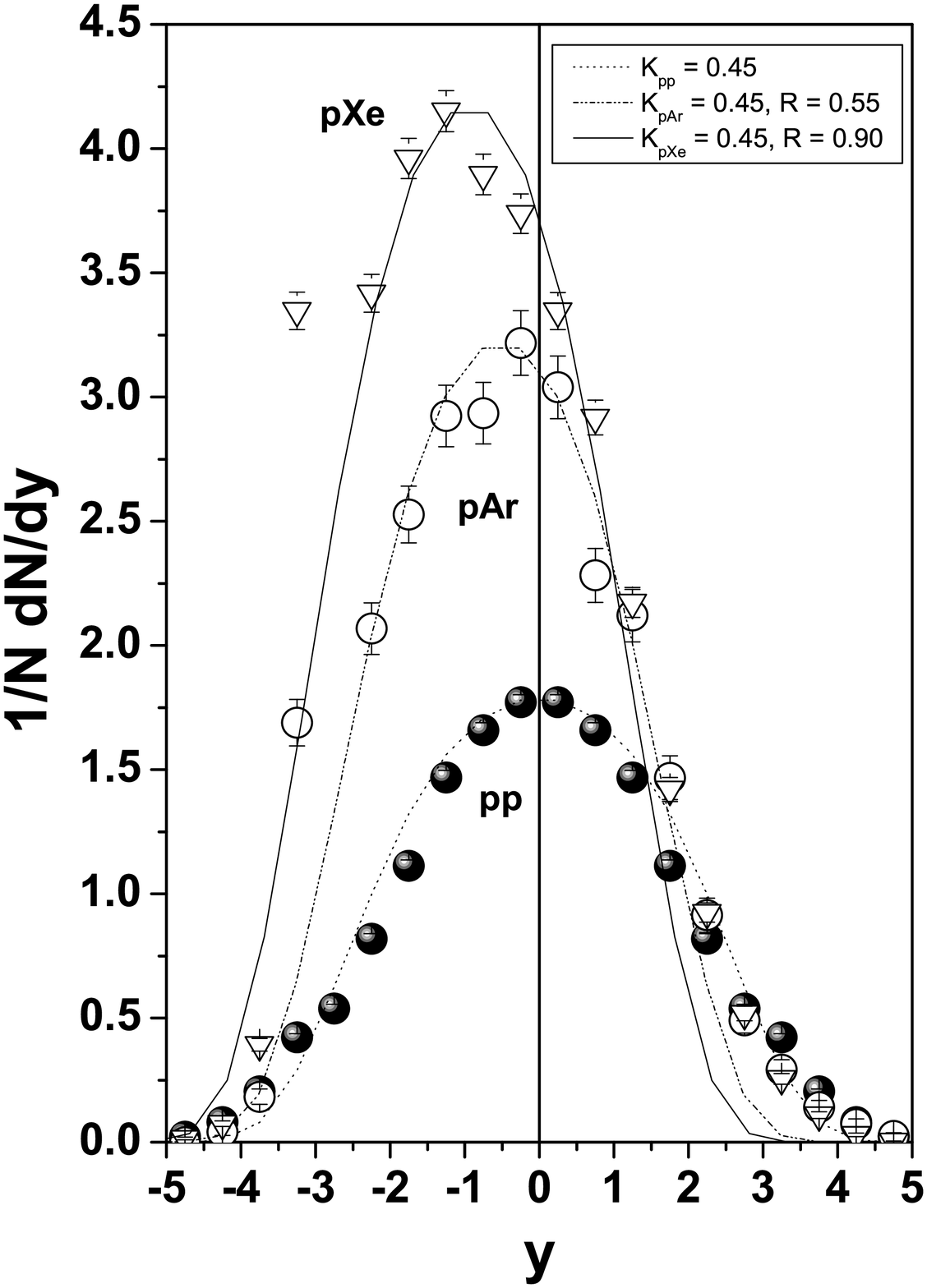}
  \end{minipage}
  \vspace{-5mm}
  \caption{ \footnotesize Example of fits to $\rm pp$, $\rm pAr$ and $\rm pXe$ data
           \protect\cite{NucData2} by means of the collective tube
           model approach as given by Eq.(\ref{eq:ftildenu}) with
           $R=K$ (left panel) and with $R \neq K$ for $pA$ collisions
           (right panel).}
 \label{fig:Fig4}
\end{figure}

In the first case we have situation similar to considered before with the following
formula for the rapidity distribution of secondaries produced in two-body $p(\nu N)$
(in the CM frame of the $NN$ system):
\begin{eqnarray}
\frac{dN_{\nu}}{dy}\, &=&\, f_{\nu}(y)\, =\, \frac{1}{Z_{\nu}}\cdot \exp\left( -
\lambda_{\nu}\mu_T \sinh y - \beta_{\nu}\mu_T \cosh y \right), \label{eq:Fnu}\\
Z_{\nu}\, &=&\, \int^{Y^{(\nu)}_{max}}_{Y^{(\nu)}_{min}}\, dy\, \exp\left( -
\lambda_{\nu}\mu_T \sinh y - \beta_{\nu}\mu_T \cosh y \right). \label{eq:Znu}
\end{eqnarray}
We have now two lagrange multipliers, $\lambda_{\nu}$ and $\beta_{\nu}$, which are
given by the corresponding energy and momentum conservation constraints:
\begin{equation}
\int^{Y^{(\nu)}_{max}}_{Y^{(\nu)}_{min}}\!\!\!\!dy\,
     e^{ - \mu_T[\lambda_{\nu}\sinh y - \beta_{\nu}\cosh y]}
\left\{
\begin{array}{c}
\cosh y \\ \sinh y
\end{array}
\right\} = \frac{Z_{\nu}}{N_{\nu} \mu_T} \left\{ \!\!\begin{array}{c} W_{\nu} = (\nu
R + K) \frac{\sqrt{s}}{2},
\\ P_{\nu} = - (\nu R - K) \frac{\sqrt{s - 4m^2}}{2}
\end{array}\!\!
\right\}. \label{eq:constr}
\end{equation}
The energy transfer from the projectile nucleon, characterized by
inelasticity $K$, is allowed to differ from energy transfers from
the nuclear nucleons (cf. Fig. \ref{fig:Fig3}b) characterized by
inelasticities $R_i$ (for simplicity we shall assume $R_i = R$ in
what follows). The invariant mass $M_{\nu}$ of hadronizing system is
now equal to
\begin{equation}
M_{\nu}\, =\, \sqrt{W^2_{\nu} - P^2_{\nu}}\, = \, \sqrt{\, KR\nu s\, +\, (\nu R -
K)^2 m^2}\quad \stackrel{R=K}{\Longrightarrow}\quad K\sqrt{s_{\nu}}  \label{eq:Mn}
\end{equation}
whereas the (longitudinal) phase space in which particles are
produced is given by:
\begin{equation}
Y^{(\nu)}_{max} = Y_{\nu m} - \delta_{\nu} \qquad {\rm and}\qquad Y^{(\nu)}_{min} = -
Y_{\nu m} - \delta_{\nu}, \label{eq:Ym}
\end{equation}
($Y_{\nu m}$ is the same as $Y_M$ before but calculated for
$M_{\nu}$) where $\delta_{\nu}$ being the rapidity shift between the
$NN$ and $(\nu N)-N$ center of mass frames:
\begin{equation}
\tanh \delta_{\nu} = - \frac{P_{\nu}}{W_{\nu}}\quad \Longrightarrow \quad
\delta_{\nu} \simeq
      \frac{1}{2}\ln \left[\frac{R}{K}\cdot\nu\right]\,
\stackrel{R=K}{\Longrightarrow}  \,  \frac{1}{2}\ln \nu \, .
                 \label{eq:delta}
\end{equation}
The apparently asymmetric form of rapidity distribution as given by
Eq.(\ref{eq:Fnu}) is, however, an artifact connected with our choice
of the reference frame. Changing variables in eqs.
(\ref{eq:Fnu})-(\ref{eq:constr}) from $y$ to $\tilde{y} = y +
\delta_{\nu}$, i.e. proceeding from the CMS of $NN$ to the CMS of
$(\nu N)-N$, one gets similar distribution as before (i.e. depending
only on one lagrange multiplier $\tilde{\beta}_{\nu}$):
\begin{eqnarray}
f_{\nu}(y)\!\! &\Rightarrow &\!\! f_{\nu}(y = \tilde{y} - \delta_{\nu})=
\frac{1}{\tilde{Z}_{\nu}}\cdot e^{ - \tilde{\beta}_{\nu} \mu_T \cosh \tilde{y}
};\quad \tilde{Z}_{\nu} = \int^{Y_{\nu m}}_{-Y_{\nu m}}\!\!\!\!\! d\tilde{y} e^{ -
\tilde{\beta}_{\nu} \mu_T \cosh \tilde{y} };\label{eq:ftildenu}\\
&&\int^{Y_{\nu m}}_{-Y_{\nu m}}\!\!\!\!\! d\tilde{y}\cosh \tilde{y}
e^{ - \tilde{\beta}_{\nu}\mu_T \cosh \tilde{y} } =
\frac{Z_{\nu}}{N_{\nu} \mu_T}\cdot M_{\nu},\label{eq:Ztilden}
\end{eqnarray}
where $M_{\nu} = M_{\nu}(K,R)$. It means that in such an approach we
always can find frame in which rapidity distributions $f_{\nu}$ are
{\it symmetric function} given by (\ref{eq:ftildenu}). In Fig.
\ref{fig:Fig4} we show examples of fits to some available data
(notice that data display clear asymmetric character which cannot be
reproduced by the method used here).

In the second case ${\rm p}A$ scattering is assumed to proceed via
sequence of the $\nu$ two-body processes, cf. Fig. \ref{fig:Fig3}c.
The resultant rapidity distribution $f_{\nu}(y)$ is then composed of
"elementary" distribution functions describing collisions of the
impinging nucleon with the subsequent nucleons in the target
nucleus:
\begin{equation}
f_{\nu}(y) = \sum_{i=1}^{\nu}f_i(y) . \label{eq:fnufni}
\end{equation}
Contrary to (\ref{eq:multi}), this formula stresses not the fact
that in the ${\rm p}A$ reaction one has $\nu +1$ participating
nucleons but that one has here $\nu$ consecutive collisions treated
as "elementary" ones. This can be visualized rewriting
(\ref{eq:multi}) as $\bar{N}_{\nu} = \bar{N_1} + (\nu -
1)\bar{N_2}/2$ where $\bar{N}_{1,2}$ are such that $2\bar{N}_1 +
(\nu -1)\bar{N}_2 = (\nu +1) \bar{N}$. We have thus $\nu$
"elementary" collisions with a possibility that the first can differ
from the remain $\nu -1$ (in general they all are different). As
result, the elementary inelasticities (i.e. inelasticities in
subsequent collisions) are not necessarily the same. In this way one
approaches as near as possible the notion of independent production
for which one demands the {\it a priori} knowledge of energy of each
reaction. Also here we shall differentiate between the fractions of
energies contributed to each $f_i$ by the nuclear nucleons,
$R_{i=1,\dots,\nu}$, and fractions of energies contributed by the
impinging nucleon at each collision, $K_{i=1,\dots,\nu}$. In this
case the general form of the energy-momenta $(e_i,p_i)$ and
$(w_i,v_i)$ flowing to the blob $M_i$ in Fig. \ref{fig:Fig3}c is
given by:
\begin{eqnarray}
e_i &=& \frac{1}{2} K_i\prod_{j=1}^{i-1}\left(1-K_j\right)\sqrt{s};\nonumber\\
p_i &=& \frac{1}{2}\left[\prod _{j=1}^{i-1}\left(1 - K_j\right)s - 4m^2\right]^{1/2}
- \frac{1}{2} \left[\prod_{j=1}^{i}\left(1-K_j\right)s - 4m^2\right]^{1/2}
\label{eq:EP}\\
w_i &=& \frac{1}{2}R_i\sqrt{s};\qquad v_i = \frac{1}{2}\sqrt{s - 4m^2} - \frac{1}{2}
\sqrt{\left(1-R_i\right)\cdot s - 4m^2} \label{eq:WV}
\end{eqnarray}
with $M^2_i = \left[ \left(e_i + w_i\right); \left(p_i + v_i\right)\right]^2$. Notice
that the real fraction of energy deposited by impinging nucleon in, say, second
collision , is equal to $K^{(2)} = K_2\cdot \left( 1 - K_1\right)$, in general
$K^{(i)} = K_i\cdot \prod_{j=1}^{i-1}\left(1 - K_j\right)$. We have now that
\begin{eqnarray}
f_i(y) &=& \frac{1}{Z_i} e^{-\beta_i\mu_T \cosh y};\label{eq:fi}\\
Z_i &=& \int^{Y_i}_{-Y_i}\!\!\!dy e^{-\beta_i\mu_T \cosh y};\qquad
\int^{Y_i}_{-Y_i}\!\!\! dy \cosh y e^{ - \beta_i\mu_T \cosh y } = \frac{Z_i}{N_i
\mu_T}\cdot M_i .  \label{eq:Mit}
\end{eqnarray}
with $Y_i$ calculated in the same way as $Y_M$ before with $M_i$ and $n_i$ replacing
$W$ and $N$, respectively. These distribution will be centered (in the CM frame of
$NN$) at
\begin{equation}
y_i \simeq \frac{1}{2} \ln \left[\frac{K_i}{R_i}\prod_{j=1}^{i-1}(1-K_j)\right]\qquad
\stackrel{R_i=K_i=K}{\Longrightarrow}\qquad \propto \frac{(i-1)}{2}\ln (1-K) ,
\label{eq:centeri}
\end{equation}
i.e. the expected shift in rapidity is now {\it linear} in the
number of collisions, not logarithmic one as in Eq.(\ref{eq:delta}).
However, for our limited values of $\nu$ this difference is
practically not visible. The last input needed to apply {\it IT} is
some estimation of multiplicities $N_i$. This is the most uncertain
point here and can be done in many model-dependent ways. Here we
have simply assumed that $N_i = a\cdot M_i^c$ (with $a$ and $c$
being free parameters) and made use the fact that
$N=\sum^{\nu}_{j=1} N_i = a\cdot \sum_{j=1}^{\nu} M_j^c $ to
eliminate parameter $a$ and to write $N_i = N/\left[1 +
\sum^{\nu}_{j \neq i}\left( M_j/M_i\right)^c\right]$, where
$c=0.556$ is obtained from reproducing the observed multiplicity $N$
(actually, logarithmic dependence $N_i = a + b\cdot \ln M_i$ would
work equally well). In order not to introduce too much freedom with
the choices of $K_i$ we were proceeding in the following way: $K_1$
was fixed by the $\nu=1$ data and then used as $K_{i=1}$, then $K_2$
was fixed by $\nu=2$ data and used as $K_{i=2}$ and so on up to
$K_4$, which was fixed by $\nu=4$ data.$^l$ On the other hand the
fractions of energy deposited by the consecutive target nucleons
participating in the collision was kept the same for all of them and
equal $R=R_{\nu}$.
\begin{figure}[ht]
  \begin{minipage}[ht]{57mm}
        \insertplot{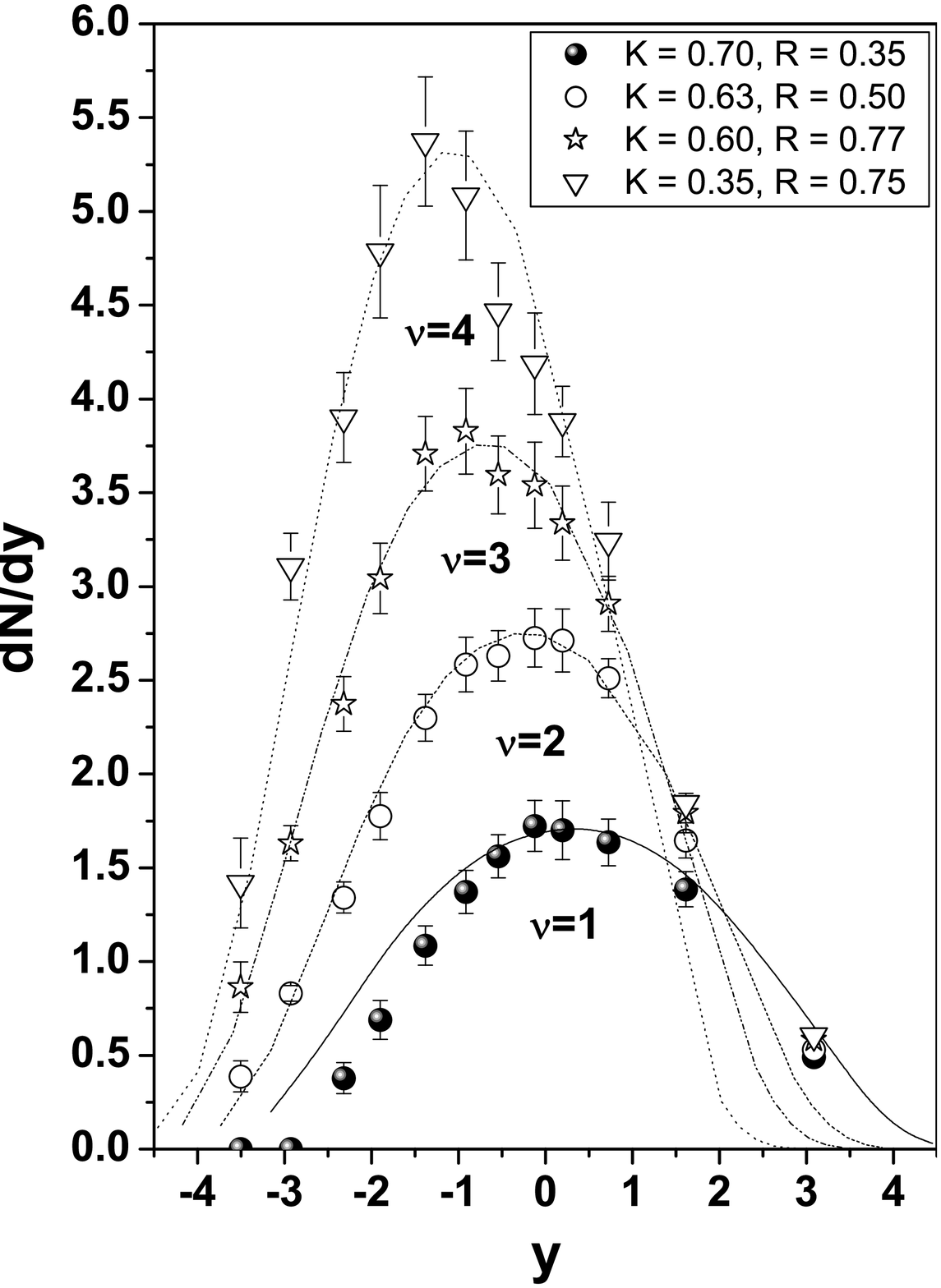}
  \end{minipage}
\hspace{5mm}
  \begin{minipage}[ht]{57mm}
        \insertplot{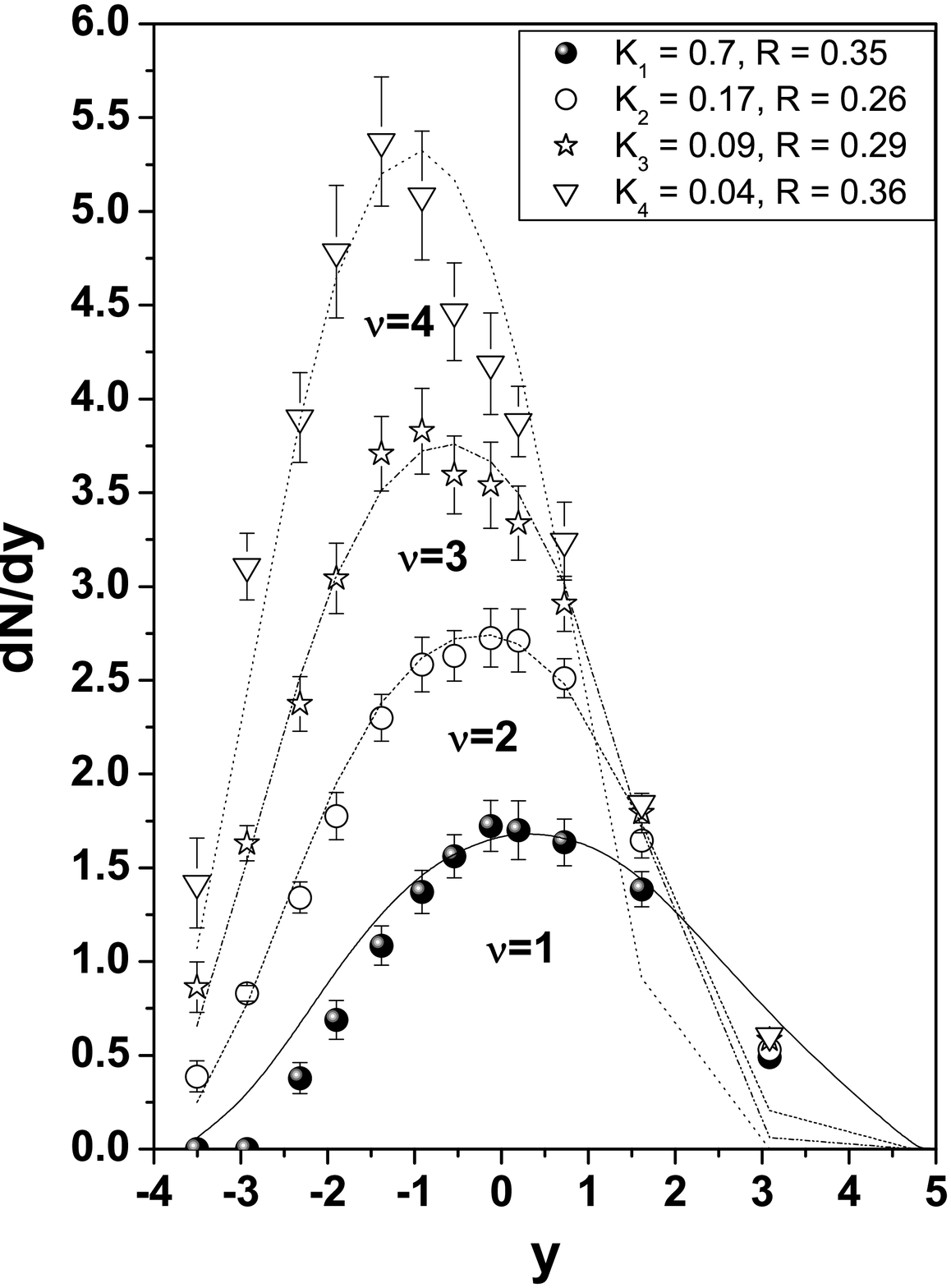}
  \end{minipage}
  \vspace{-3mm}
  \caption{ \footnotesize Left panel: Comparison with $pA$ data for
           $\nu$ \cite{NucData1} using tube model (Eq. (\ref{eq:ftildenu})
           and Fig. \ref{fig:Fig3}a) for $K\neq R$. Notice
           that data for $\nu=1$ are not compatible with data for
           $pp$ collisions from \cite{NucData2} presented in Fig.
           \ref{fig:Fig3}. Right panel: the same but using
           sequential model (Eq.(\ref{eq:fnufni}) and Fig. \ref{fig:Fig3}c).
           $K_i$ is the fraction of the actual projectile
           energy in the $i-th$ collision deposited in central region.
           The fractions of energy deposited by target nucleons is
           kept the same and equal $R=R_{\nu}$.}
 \label{fig:Fig5}
\end{figure}

Notice that in the sequential model we have clearly smaller
consecutive energy transfers (as given by $K_{i>1}$) from the
nucleon traversing the nucleus (this follows observation made in
\cite{KinWW}). On the other hand, both approaches lead to
essentially the same results. However, only sequential model is
potentially able to reproduce the visible asymmetry seen in data for
larger $\nu$ because "elementary" $f_i$ enter with different weights
and cover slightly different regions of phase space. However, at
present energies the effect is not as large as expected and
sequential model leads to essentially the same results as tube
model.$^m$ To account for the shift in the position of maximum of
rapidity distribution in the tube model we have to allow for $K\neq
R$, i.e. for different energy depositions from each of the
projectiles. However, even then we cannot produce the asymmetry of
rapidity distribution seen in data, in particular we are not able to
correctly describe the part of distribution connected with the
impinging proton. It should be stressed that using in this case the
nonextensive approach {\it ITq} would not help because also in this
case the obtained distributions are symmetric in the rest frame of
the hadronizing system. There is clearly difference between the
impinging propton and nuclear parts of the phase space. We conclude
therefore that, according to the rules of information theory
approach, data \cite{NucData1,NucData2} still bear some additional
information, not identified so far.

\section*{Summary and Conclusions}

We have demonstrated that large amount of data on multiparticle
production can be quite adequately described by using tools from
information theory, especially when allowing for its nonextensive
realization based on Tsallis entropy. We have argued that
nonextensivity parameter $q$ entering here can, in addition to the
temperature parameter $T$ of the usual statistical approaches,
provide us valuable information on dynamical fluctuations present in
the hadronizing systems. Such information can be very useful when
searching for phase transition phenomena, which should be
accompanied by some specific fluctuations of non-monotonic character
\cite{fluc}.

As concerns the asymmetric collisions example of ${\rm p}A$ data, we
have demonstrated that they contain additional information to the
usual one used in Section \ref{RESYM}. The immediate candidates are
the rescattering of produced secondaries in the nucleus (effectively
increasing values of $\mu_T$ and making it $y-$dependent and in this
way limiting rapidity space available on the nuclear side) or
diffraction dissociation part of the production (which can take
different shape than in the $\rm pp$ collisions), but they can be
other possibilities as well \cite{MaxEntq}.

Finally let us comment shortly on the inelasticities in ${\rm p}A$
collisions obtained. In the tube model they are, in general,
increasing (or, at least, non-decreasing) with $A$ or with $\nu$,
cf. Figs. \ref{fig:Fig3}, \ref{fig:Fig4} and \ref{fig:Fig5}. Notice
that for $K\neq R$ case the effective inelasticity, which is of
interest for any statistical model approach (i.e. the part of the
total energy available for production of secondaries in the central
region of reaction) is $K_{eff}=\sqrt{K\cdot R}$. In sequential
model this inelasticity is clearly decreasing with the consecutive
collisions (in agreement with what was found in \cite{KinWW}).
However, the more precise statement could be only done with a data
taken at much higher energies. Notice that we did not present fits
to ${\rm pAr}$ and ${\rm pXe}$ data \cite{NucData2} in this case
because we would have to introduce here another piece of additional
information represented by distribution of number of collisions,
$P(\nu)$, which is model dependent quantity. The high energy
counterpart of data \cite{NucData1} would be therefore most welcome.
The point is, however, that in addition one should also have data
taken for $pn$ inelastic collisions (possibly with the help of
deuterons) as it is possible that at least part of the inconsistency
between $pp$ data from \cite{NucData2} (cf., Fig. \ref{fig:Fig4})
and $\nu=1$ data from \cite{NucData1} in Fig. \ref{fig:Fig5} is
because the later contain also contribution from $\rm pn$
reactions.$^n$

We close with the statement that results presented here are very
encouraging and call for further systematic effort to describe existing
data in terms of $(T,q)$ for different configurations and energies in
order to find possible regularities in their system and energy
dependencies and possible correlations between them.

\section*{Acknowledgments}
Two of us (OU and GW) are grateful for support from the Hungarian Academy
of Science and for the warm hospitality extended to them by organizers of
the {\it $4^{th}$ Budapest Winter School on Heavy Ion Physics}, Dec. 1-3,
2004, Budapest, Hungary. Partial support of the Polish State Committee
for Scientific Research (KBN) (grant 621/E-78/SPUB/CERN/P-03/DZ4/99 (GW))
is acknowledged.

\section*{Notes}
\begin{notes}
\item[a] The complete list of references concerning {\it
IT} relevant to our discussion and providing necessary background
can be found in \cite{MaxEnt,MaxEntq}. The value of Boltzmann
constant is set to unity, $k=1$.
\item[b] Notice that using the entropic
measure $S\, =\, \sum_i\left[p_i \ln p_i\, \mp\, \left(1\, \pm\,
p_i\right) \ln \left(1\, \pm\, p_i \right) \right]$ (which, however,
has nothing to do with {\it IT}) would result instead in
Bose-Einstein and Fermi-Dirac formulas: $p_i\, =\, (1/Z)\cdot
\left[\exp[\beta(\varepsilon_i - \mu)] \mp 1\right]^{-1}$, where
$\beta$ and $\mu$ are obtained solving two constraint equations
given, respectively, by energy and number of particles conservation
\cite{TMT}. It must be also stressed that the final functional form
of $p_i$ depends also on the functional form of the constraint
functions $R_k (x_i)$. For example, $R(x) \propto \ln (x)$ and $\ln
(1-x)$ type constrains lead to $p_i \propto
x_i^{\alpha}(1-x_i)^\beta$ distributions.
\item[c] It is straightforward to check that the Shannon entropy decreases from the
most broad geometrical distribution towards the most narrow
Poissonian distribution.
\item[d] The most drastic situation was with the multi-Regge model in which,
in addition to the basic model assumptions, two purely
phenomenological ingredients have been introduced in order to get
agreement with experiment: $i)$ energy $s$ was used in the scaled
$(s/s_0)$ form (with $s_0$ being free parameter, this works the same
way as inelasticity) and $ii)$ the so-called "residual function"
factor $e^{\beta\cdot t}$ was postulated ($t=-(p_i-p_j)^2$ and
$\beta$ being a free parameter) to cut the transverse part of the
phase space. Therefore $s_0$ and $\beta$ were {\it the only
relevant} parameters.
\item[e] For our purpose this is sufficient and there is no need to use more
sophisticated approach exploring escort probabilities formalism, see
\cite{Kq} (for most recent discussion of different constraints and
their meaning see \cite{HS} and references therein).
\item[f] Strictly speaking in \cite{WW} it was shown only for
fluctuations of $1/\Lambda$ given by gamma distribution. However, it
was soon after generalized to other form of fluctuations and the
word {\it superstatistics} has been coined to describe this new
phenomenon \cite{CB}. Another generalization of this idea can be
found in \cite{Biro}.
\item[g] Notice that the $y$ space is limited to $ y\in(-Y_M, Y_M)$ where (with
$\mu_T=\sqrt{m^2 +\langle p_T\rangle^2}$ being the mean transverse
mass and $M' =M - (N-2)\mu_T$ accessible kinetic energy) the limits
are given by $$Y_M = \ln \left\{ M'/(2\mu_T) \left[1+\left( 1-
4\mu_T^2/M'^2 \right)^{1/2}\right]\right\}$$.
\item[h] It was jus a coincidence that at ISR
energies this condition was satisfied. But because such a behavior
of $N$ as function of energy was only some transient phenomenon,
there will never be scaling of this type at higher energies,
notwithstanding all opinions to the contrary heard from time to
time.
\item[i] In \cite{NuCim} we
have fitted data on $\rm p\bar{p}$ with $q<1$ being the only
parameter and $M=\sqrt{s}$. The $q<1$ was cutting off the available
phase space playing effectively the role of inelasticity $K$. This
result has provided the cosmic ray physicist community justification
of the empirical formula they used, namely that $f(x) \propto
(1-a\cdot x)^n$, where $x$ is Feynman variable, $x=2E/\sqrt{s}$, and
where $a$ and $n$ are free parameters. They turned out to be both
given by the parameter $q$ only \cite{NuCim}.
\item[j] In what follows only
interactions resulting in the production of particles in the central
region of reaction are of interest to us, elastic and diffractive
dissociation collisions will not be considered.
\item[k] Although this
regularity has been observed only for charged secondaries, we shall
assume here its validity both for the total number of produced
particles as well as for the fixed number of collisions $\nu$.
\item[l] Notice that such procedure is possible only under the assumption of
independent collisions.
\item[m]We would like to stress at this
point that our approach to ${\rm p}A$ using {\it IT} concepts and
tube model differs from that of \cite{LY} because we have only one
parameter, inelasticity $K=R$, with both the normalization, shift of
the momenta and "partition temperature" being fixed by it, whereas
in \cite{LY} they are all free parameters.
\item[n] We shall not pursued
further this problem, which in our opinion can be investigated in
the spirit of {\it IT} only when the same experiment will provide
data both for different and well defined values of $\nu$ and for
$\langle \nu\rangle$, i.e. averaged over different $\nu$ (such
possibility was apparently under consideration by NA49 Collaboration
\cite{NA49pA}). We are also aware of potentially interesting new
data from RHIC on ${\rm d}Au$ collisions \cite{dA} and recent
attempts of their description \cite{BC} and we plan to address this
issue using {\it IT} approach elsewhere.
\end{notes}

\vfill\eject

\begin{thebibliography}{99}
\bibitem{Stat} Cf., for example: F.~Becattini, {\sl Nucl. Phys.} {\bf A702}
               (2002) 336; W.~Broniowski and W.~Florkowski,
               {\sl Acta Phys. Polon.} {\bf B35} (2004) 779, and references
               therein.

\bibitem{LVH} L.~Van Hove, {\sl Z. Phys.} {\bf C21} (1985) 93; ibid.
              {\bf C27} (1985) 135.

\bibitem{Alm} M.P.~Almeida, {\sl Physica} {\bf A325} (2003) 426.

\bibitem{MaxEnt} G.~Wilk and Z.~W\l odarczyk, {\sl Phys. Rev.} {\bf D43}
                 (1991) 794.

\bibitem{MaxEntq} F.S.~Navarra, O.V.~Utyuzh, G.~Wilk and
                  Z.~W\l odarczyk, {\sl Physica} {\bf A340} (2004) 467.

\bibitem{TMT} A.M.~Teweldeberhan, H.G.~Miller and R.~Tegen, {\sl Int. J.
              Mod. Phys.} {\bf E12} (2003) 395.

\bibitem{NBin} P.~Carruthers and C.C.~Shih, {\sl Int. J. Mod. Phys.} {\bf A4}
               (1989) 5587.

\bibitem{Chao} Y.-A.~Chao, {\sl Nucl. Phys.} {\bf B40} (1972) 475.

\bibitem{Gauss} R.S.~Johal, A.~Planes and E.~Vives, {\sl Phys. Rev.}
                {\bf E68} (2003) 056113.

\bibitem{ENT} F.~Tops\oe, {\sl Physica} {\bf A340} (2004) 11.

\bibitem{CT} C.~Tsallis, in {\it Nonextensive Statistical Mechanics
             and its Applications}, S.Abe and Y.Okamoto, eds.,
             {\it Lecture Notes in Physics} LPN560, Springer (2000), in
             {\sl Physica} {\bf A340} (2004) 1; {\sl Physica}
             {\bf A344} (2004) 718, and references therein.

\bibitem{Kq} F.S.~Navarra, O.V.~Utyuzh, G.~Wilk and Z.~W\l odarczyk,
             {\sl Phys. Rev.} {\bf D67} (2003) 114002.

\bibitem{HS} H.~Suyari, Refined formalism of the maximum entropy principle in
             Tsallis statistics, cond-mat/0502298.

\bibitem{Biya} See, for example, M.~Biyajima, M.~Kaneyama, T.~Mizoguchi and G.~Wilk,
               {\sl Eur. Phys. J.} {\bf C40} (2005) 243 and references therein.

\bibitem{WW} G.~Wilk and Z.~W\l odarczyk, {\sl Phys. Rev. Lett.} {\bf 84}
             (2000) 2770; {\sl Chaos, Solitons, Fractals} {\bf 13}
             (2002) 581; {\sl Physica} {\bf A305} (2002) 227.

\bibitem{CB} C.~Beck and E.G.D.~Cohen, {\sl Physica} {\bf A322} (2003) 267.

\bibitem{Biro} T.S.~Bir\'o and A.~Jakov\'ac, {\sl Phys. Rev. Lett.} {\bf
               94} (2005) 132302.

\bibitem{NuCim} F.S.~Navarra, O.V.~Utyuzh, G.~Wilk and
                Z.~W\l odarczyk, {\sl Nuvo Cim.} {\bf C24} (2001) 725.

\bibitem{Cern} M.~Rybczy\'nski, Z.~W\l odarczyk and G.~Wilk, {\sl Nucl.
               Phys.} ({\sl Proc. Suppl.}) {\bf B122} (2003) 325.

\bibitem{Trends} F.S.~Navarra, O.V.~Utyuzh, G.~Wilk and
                 Z.~W\l odarczyk, {\sl Physica} {\bf A344} (2004) 568.

\bibitem{JPG} O.V.~Utyuzh, G.~Wilk and Z.~W\l odarczyk, {\sl J. Phys.} {\bf
              G26} (2000) L39.

\bibitem{Nukleon} F.S.~Navarra, O.V.~Utyuzh, G.~Wilk and
                  Z.~W\l odarczyk, {\sl Nukleonika} {\bf 49}
                  (2004) ({\sl Supplement}) S19.

\bibitem{fluc} M.~Rybczy\'nski, Z.~W\l odarczyk and G.~Wilk,
               {\sl Acta Phys. Polon.} {\bf B35} (2004) 819.

\bibitem{CY} T.T.~Chou and C.N.~Yang, {\sl Phys. Rev. Lett} {\bf 54} (1985)
             510 and {\sl Phys. Rev.} {\bf D32} (1985) 1692.

\bibitem{NB} C.~Geich-Gimbel, {\sl Int. J. Mod. Phys.} {\bf A4} (1989)
             1527.

\bibitem{NucData1} W.~Busza, {\sl Acta Phys. Polon.} {\bf B8} (1977) 333;
                   C.~Halliwell et al., {\sl Phys. Rev. Lett.} {\bf
                   39} (1977) 1499.

\bibitem{NucData2} C.~De Marzo et al., {\sl Phys. Rev.} {\bf D26}
                  (1982) 1019 and {\bf D29} (1984) 2476.

\bibitem{LY} T.S.~Li and K.~Young, {\sl Phys. Rev.} {\bf 34} (1986) 142.

\bibitem{gray} B.~Andersson, I.~Otterlund and E.~Stenlund, {\sl Phys.
               Lett.} {\bf B73} (1978) 343.

\bibitem{KinWW} G.~Wilk and Z.~W\l odarczyk, {\sl Phys. Rev.} {\bf D59}
                (1999) 014025 and C.R.A.~Augusto et al. {\sl Phys.
                Rev.} {\bf D61} (2000) 012003.

\bibitem{NA49pA} A.~Rybicki, {\sl Acta Phys. Polon.} {\bf B33} (2002) 1483,
                 and references therein.

\bibitem{dA} B.B.~Back {\it et al.} (PHOBOS Collab.), presented by R.~Nouicer at {\it QM
             2004}; R.~Nouicer {\it et al.}, {\sl J. Phys.} {\bf G30} (2004) S1133
             [nucl-ex/0403033]; I.~Arsene {\it et al.} (BRAHMS Collab.),
             nucl-exp/0401025.

\bibitem{BC} A.~Bia\l as and W.~Czy\.z, {\sl Acta Phys. Polon} {\bf B36} (2005) 905, and references
             therein.

\end{thebibliography}
\end{document}